\documentclass[journal=jacsat,manuscript=article,layout=twocolumn]{achemso}

\usepackage[version=3]{mhchem} % Formula subscripts using \ce{}

\author{Konstantin V. Zakharchenko}
\affiliation{Nordita, KTH Royal Institute of Technology and Stockholm University,
Roslagstullsbacken 23, SE-106 91 Stockholm, Sweden}
\email{kons@nordita.org}
\author{A.V. Balatsky}
\email{avb@nordita.org}
\affiliation{Nordita, KTH Royal Institute of Technology and Stockholm University,
Roslagstullsbacken 23, SE-106 91 Stockholm, Sweden}
\altaffiliation{Theoretical Division, Center for Integrated Nanotechnologies, Los Alamos National Laboratory, Los Alamos, NM 87545, USA.}
\title[\texttt{achemso} demonstration]
{Controlled healing of graphene nanopore}

\begin{document}
%%%%%%%%%%%%%%%%%%%%%%%%%%%%%%%%%%%%%%%%%%%%%%%%%%%%%%%%%%%%%%%%%%%%%
%% The manuscript does not need to include \maketitle, which is
%% executed automatically.  The document should begin with an
%% abstract, if appropriate.  If one is given and should not be, the
%% contents will be gobbled.
%%%%%%%%%%%%%%%%%%%%%%%%%%%%%%%%%%%%%%%%%%%%%%%%%%%%%%%%%%%%%%%%%%%%%
\begin{abstract}
  Nanopores -- nanometer-size holes are very promising devices for many applications: DNA sequencing, sensory, biosensoring and molecular detectors, catalysis and water desalination. These applications require accurate control over nanopores size. We report computer simulation studies of regrowth and healing of graphene nanopores of different sizes ranging from $30$ to $5$~\AA. We study mechanism, speed of nanopores regrowth and structure of ``healed'' areas in the wide range of temperatures. We report existence of at least two distinct healing mechanisms, one so called edge attachment where carbons are attached to the edges of graphene sheet and another mechanism that involves atom insertion directly into a sheet of graphene even in the absence of the edges.

  These findings point a significantly more complicated pathways for graphene annealing. They also provide an important enabling step in development of graphene based devices for numerous nanotechnology applications.
\end{abstract}

%%%%%%%%%%%%%%%%%%%%%%%%%%%%%%%%%%%%%%%%%%%%%%%%%%%%%%%%%%%%%%%%%%%%%
%% To do part
%% 1. Check T_critical for insertion healing and find energy barrier.
%% 2. Determine healing rate for insertion as F(T).
%% 3. Check max-paper-size and max-num-figs.
%% 4. Correct addresse and e-mails.
%% 5. Temperature of flow
%% 6. Figures 1 and 3 can be merged.
%%%%%%%%%%%%%%%%%%%%%%%%%%%%%%%%%%%%%%%%%%%%%%%%%%%%%%%%%%%%%%%%%%%%%

\section{Introduction}
Recent discovery of graphene revolutionized material science and attracted close attention of the scientific community~\cite{ANIE:ANIE201101174, ANIE:ANIE201101502}. Graphene nanopores have emerged as a powerful tool for many applications: it has been shown, that graphene nanopores can be used for DNA sequencing~\cite{DNAGraphene2010}, single molecule detections~\cite{DetectMolNanopore2005}, biosensing~\cite{doi:10.1021/nn301368z, Pumera2011308} and water desalination~\cite{GrapheneDesalination2012} to name a few.

For graphene to be a material of choice for potential applications it is often desirable to produce nanostructures with the prescribed size of the pores. Preparation of the pores is a complicated process and its investigation is in its infancy.

For the application purposes there is a significant demand for samples with controlled sizes of graphene nanopores and structural integrity of the lattice around~\cite{doi:10.1021/nl200369r, Artyukhov18092012}. Any modern technology, such as electron beam drilling~\cite{fabsuspend2012}, limits the minimum diameters of nanopores to 2-5~nm~\cite{garaj2010, fischbein:113107}, while for example water desalination requires nanopores of maximum size 5.5~\AA~\cite{GrapheneDesalination2012}. In addition, damage from the electron beam yields undesired amorphization of the crystalline lattice structure around~\cite{Terdalkar20083908}, which changes electronic properties of graphene nanodevice.  Frequently used method of graphene healing is annealing the sample in the ambient conditions at high temperatures with the supply of extra carbon atoms~\cite{garaj2010}. The atoms then attach to the graphene structure and regrow the pores. While this process is known empirically we find it useful to provide a more detailed microscopic description of the regrowth and healing process.

In this article we investigate the details of graphene sheet growth. We study regrowth and healing of graphene nanopores by means of molecular Monte Carlo simulations in various conditions in the \textit{NPT} ensembles (constant number of particles \textit{N}, constant pressure \textit{P} and constant temperature \textit{T}) with periodic boundary conditions in the plane for samples of N = 4032 before the creation of nanopore.

We found two distinct healing mechanisms of the nanopore. The first involves reconstruction of the lattice at the open edges of the pore, where carbon atoms approach and attach at the edge. This mechanism is active as long the edge, such as nanopore exists and at all range of temperatures. The second mechanism works when the temperature is above critical $T_c = 900$~K. This mechanism is best described as the insertion of carbon atoms directly into the crystalline structure of graphene. These insertions results in the formation of the defects and slow drift of these defects towards the pore. The second mechanism is present without any edges and therefore operates even in pristine graphene sheet. Both mechanisms are described below in details.

Outline of the paper is as follows: we describe first two different mechanisms of the pore healing in details in section {\em Healing Mechanisms}. Subsequently, we discuss structure of healed area and sample in general, and rate of healing as a function of temperature in {\em Structure of the healed nanopore}. Finally, we conclude with summary and future challenges. Details of our approach can be found in the {\em Methods} section in the supplement.

\section{Healing mechanisms}

We start with the open edge healing of the nanopore and in separate subsection describe the alternative atom insertion mechanism. To simulate the healing of the nanopore we inject carbon atoms in series of 32 or 18 in the free space above and below free standing graphene sample. Before starting the simulation, we ensured that neither two atoms are close enough to become immediately bonded to the graphene sample or between each other. We should point that we focus on the growth of a free standing graphene. The growth of graphene on the substrate is  a more complicated matter and deserves a separate discussion.

\subsection{Edge healing}

\begin{figure}
  \centering
\includegraphics[clip=true,width=0.9\linewidth]{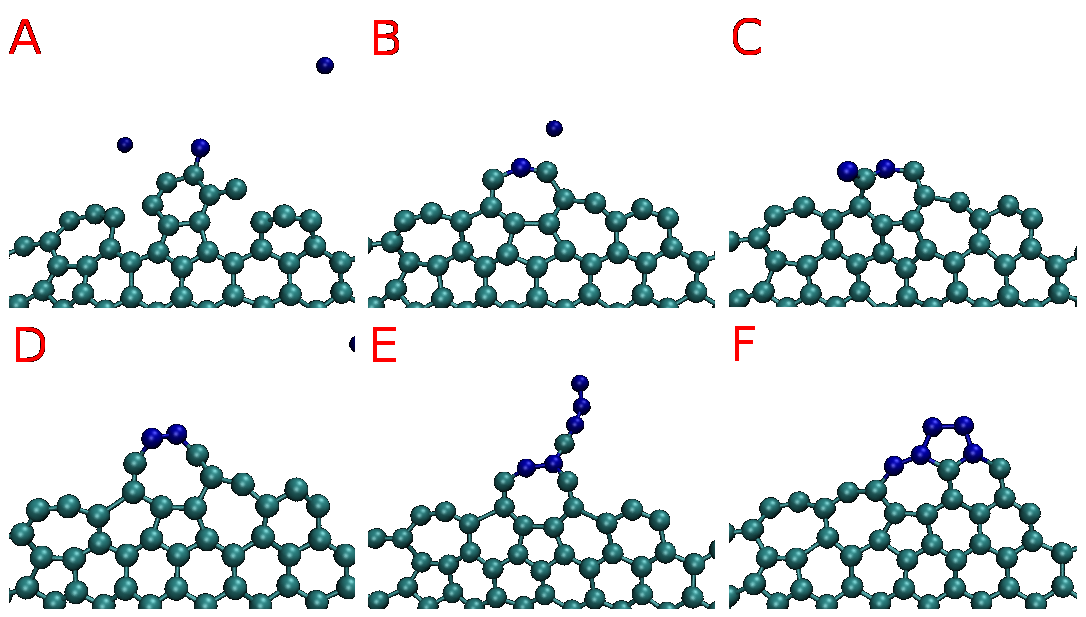}
\caption{Edge healing mechanism: A) free carbon atom attaches to the open edge; B) edge reconstruction - new ring of carbon atoms is formed; C) second atom attached in the vicinity of the previous one; D) new edge reconstruction; E) formation of a chain of carbon atoms; F) reconstruction of the edge and formation of new 5 and 7 rings using atoms of the chain.}
\label{fig1a}
\end{figure}

More than half of the atoms at the edge have less than 3 neighbours, which results in the presence of dangling bonds. When carbon atom appears in the close vicinity of the edge of the nanopore, it bonds to the edge (see Fig.~\ref{fig1a}A). Edge reconstruction proceeds by forming new ring of carbon atoms as shown in~Fig.~\ref{fig1a}B or a dangling atom. At the next stage, it is possible, although does not have to happen immediately afterwards, that new carbon atom will be attached in the close vicinity of the previous one (see Fig.~\ref{fig1a}C). Figure~\ref{fig1a}D shows the result of the reconstruction of the edge. This process repeats many times, thus reconstructing the structure of graphene lattice, healing the nanopore and reducing its size~\cite{JPC2010689, Artyukhov18092012, PRB2012205448}. Another less frequent scenario we observed for the edge healing, is the formation of a chain of carbon atoms attached to the edge of the nanopore like one can see in Fig.~\ref{fig1a}E. The longer the chain the less stable it is. After a short time (from one to ten thousand of Monte Carlo steps depending on the temperature and chain length) such chains become a part of the structure of graphene sample. In this case (Fig.~\ref{fig1a}E) we can se the formation of new 5 and 7 rings (defected structure), the structure developed is ``half'' of Stone-Wales defect well known for graphitic structures.

This mechanism works as long the open edge is present, or in other words until the nanopore exists. The fact that new atoms attaches to the dangling bonds, means the absence of the energy barrier and thus healing rate will be practically unaffected by the temperature.

\begin{figure}
  \centering
\includegraphics[clip=true,width=0.9\linewidth]{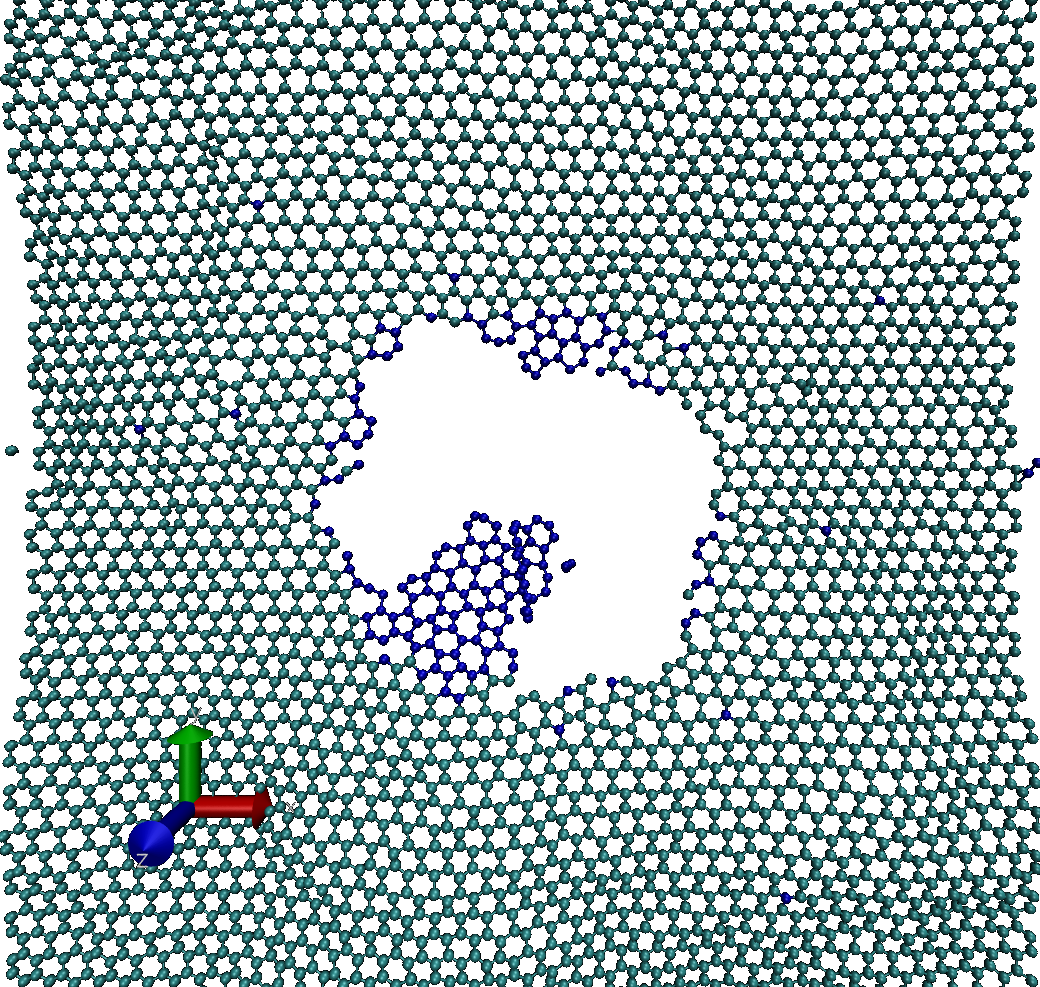}
\caption{Part of the asymmetrically healed nanopore with ``finger-like'' protrusion in healed area. Initial nanopore radius 20 \AA, healing temperature 2100~K.}
\label{fig1f}
\end{figure}

As mentioned, the formation of chain events are (Fig.~\ref{fig1a}E)  relatively rare. In the most cases, atoms attach to the edge one by one, with reconstruction of the edge after each event.  In several cases we observed that small chain can reach 10-15 atoms in length and become a ``center of condensation'' of other carbons and starting point for the reconstruction of relatively big area of graphene. It results in the asymmetry of the shape of the healed nanopore, see Fig.~\ref{fig1f}. The structure of the reconstructed lattice is not ideal, albeit less disordered than amorphous. The type and number of defects depend on the temperature of the healing and will be discussed later.

 The analogous ``finger-like'' instability of a free surface is common phenomenon in much more complex biological systems. For example, the collective motion of epithelial cells during wound healing occurs via fingering destabilisation of the border~\cite{wndpnas} and results in a similar structure as in Fig.~\ref{fig1f}. We believe, that such similarities would be more noticeable for nanopores of bigger sizes.

\subsection{Atoms insertion healing}

\begin{figure}
  \centering
\includegraphics[clip=true,width=0.9\linewidth]{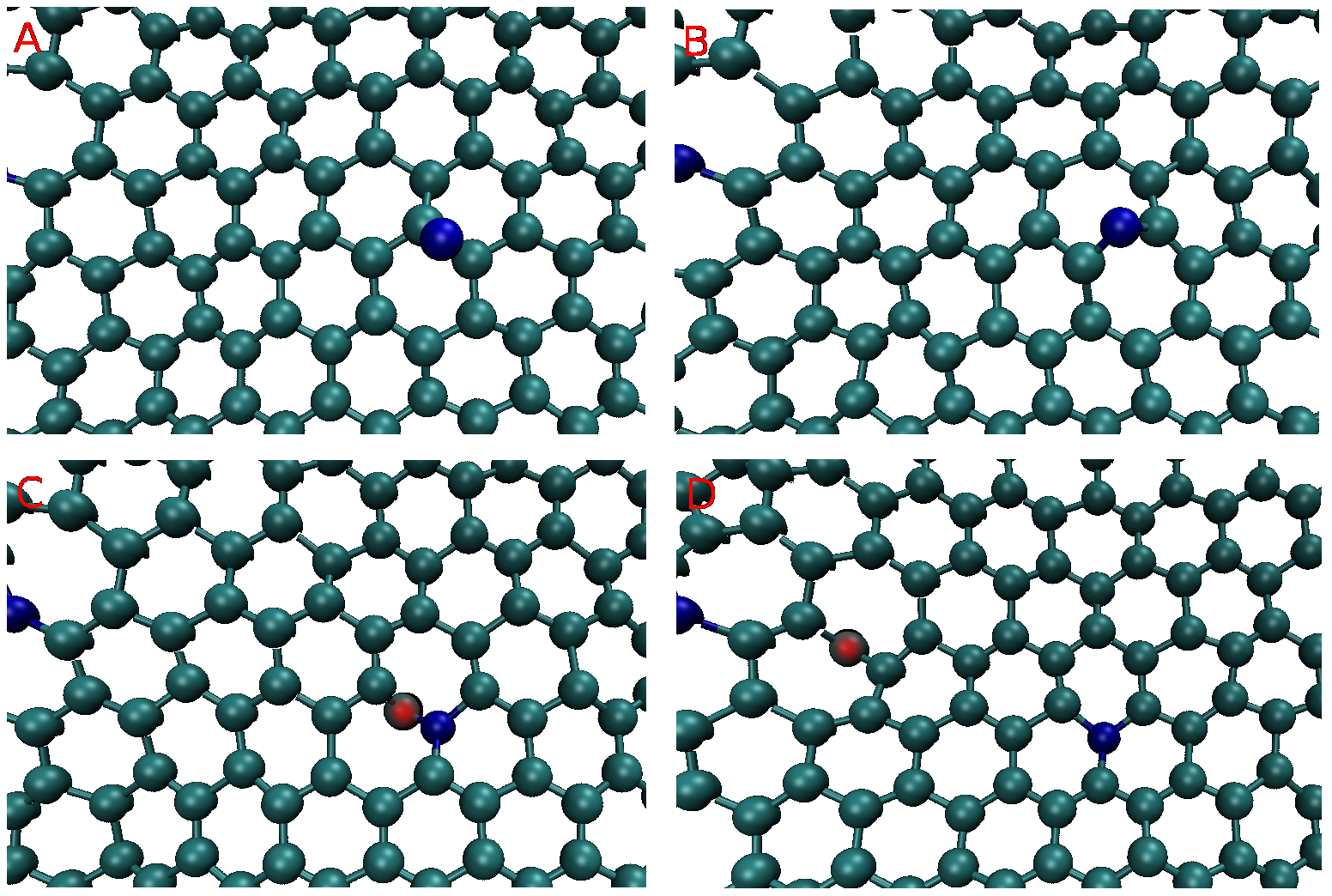}
\caption{Surface healing mechanism: A) free carbon atom binds to the graphene surface; B) formation of topological defect which consist of two 7-rings; C) newly added atom ``displaces'' original atom (marked with red dot), defect starts to move; D) position of defect (marked with red dot) after 15000 steps, defect moved $\approx 10$ \AA~left-up from the initial position towards the nanopore.}
\label{fig2a}
\end{figure}

The second healing mechanism we found is the atom insertions when atom inserts directly into a graphene plane. When carbon atom appears in the close vicinity of graphene plane but away from the edge of the nanopore, this atom will bond to the surface of graphene sample (Fig.~\ref{fig2a}A). This configuration tends to evolve into energetically more favorable (energy gain $\Delta E=1.2$~eV), if the temperature is hight enough to overcome energy barrier. Previously attached atom become bonded to two adjacent carbon atoms in graphene plane, forming two 7-rings (instead of two 6-rings) and thus topological defect as shown in (Fig.~\ref{fig2a}B). We found that energy barrier separating configurations in Figs.~\ref{fig2a}A and~\ref{fig2a}B is $\Delta E=1.85$~eV, which is relatively high. As a result of this barrier, the transition between states practically does not occur if temperature in below $T_c \approx 900$~K. After defect is formed, it can easily move inside graphene plane in any direction like shown in Fig.~\ref{fig2a}C. The motion of defect is similar to the Brownian motion of a particle, see Supplementary.

Radius-vector pointing to the nanopore from the place where atom was initially attached indicates the general direction of the motion of the defect. Thus, excessive carbon atoms displaced towards the edge of the nanopore. Next figure (Fig.~\ref{fig2a}D) shows atomic configuration of the same area after 15 thousand MC steps. Here we can clearly see that: i) at the initial position where new carbon has been attached there is a perfect lattice; ii) defect moved approximately 10~\AA~to the left and up, towards the nanopore.

In reality these two healing mechanisms works simultaneously. The relative importance of each of them depends on the size of the pore and temperature. The smaller the size of the nanopore - the greater is the role played by the surface healing. Moreover, even when the nanopore is nearly healed, atoms still binds to the surface, introducing defects in the crystalline structure.

\section{Healing rate}

Healing rate of the nanopore is directly proportional to the supply rate of carbon atoms in the sample. In addition, the attachment rate will depend on temperature. We will present the result assuming constant supply rate in all of our simulations.

To determine the quantitative behaviour of the healing rate, we use increase of the number of carbon atoms with respect to the number of Monte Carlo steps, or Monte Carlo time $\partial N/\partial t \propto \partial S/\partial t$, where $S$ is healed area. As we described above, at the temperatures below $T=900$~K healing rate drastically decreases, because atom insertions mechanism ceases to work and carbon atoms attached to the graphene surface practically do not get involved into healing. Thus, we used healing rate at $T=900$~K as a unit of measure, and express healing rates at other temperatures with respect to the healing rate at $T=900$~K.

\begin{figure}
  \centering
\includegraphics[clip=true,width=0.9\linewidth]{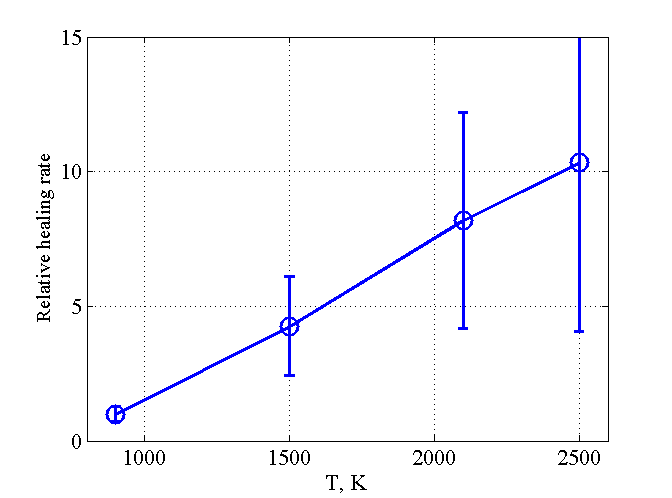}
\caption{Relative healing  rate $\partial S/\partial t$ as function of temperature.}
\label{fig3}
\end{figure}

Figure~\ref{fig3} shows relative average healing rate for different simulation temperatures. For every temperature presented result is an average of 20 to 40 simulations for every studied nanopore diameter. As we can see healing rate is linearly proportional to the temperature and increasing with the healing temperature. We also found that healing rate is practically independent on nanopore size. Taking into account that, $\partial S/\partial t \propto R \partial R/\partial t$, and thus $\partial R/\partial t \propto 1/R \partial S/\partial t \propto 1/\sqrt(S)\partial S/\partial t$, where $R\propto \sqrt S$ is characteristic size of the pore. Constant healing rate $\partial S/\partial t$ meant that radial healing $\partial R/\partial t$ rate increases during healing, which in turn requires fine control in the final stage or if target size of the pore is small.

\section{Structure of the healed nanopore}

We now discuss the structure of healed area and how it depends on the healing temperature. Figure~\ref{fig1f} shows reconstructed structure of graphene sample. Easy to notice that reconstructed area is not perfect and consists not only of hexagons, but also of pentagons, heptagons and octagons. In the perfect graphene sample carbon atoms arranged in a honeycomb lattice and each atom is connected to three its nearest neighbours. Therefore, imperfection can be described as a number of non-hexagons with respect to a number of hexagons in the structure.

\begin{figure}
  \centering
\includegraphics[clip=true,width=0.9\linewidth]{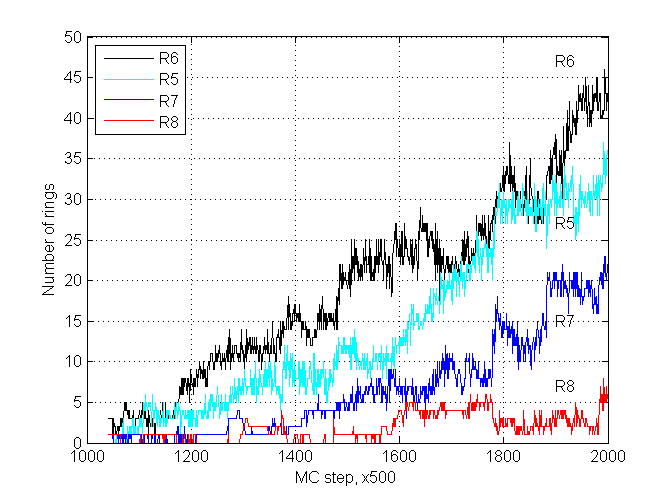}
\caption{Structure of healed area, sample size 100*100 \AA, initial nanopore radius 20 \AA.}
\label{fig4}
\end{figure}

Figure~\ref{fig4} shows number of ``rings'' of different types in the healed area for the sample with nanopore of $R=20$~\AA~which was healed at $T=1500$~K. For the healing temperature of 900 and 1500 K, the number of hexagons (``R6'' rings) is a approximately 40\% of the total number of structural rings in the healed area, the number of heptagons (``R7'') and octagons (``R8'') together is equal to approximately 15-20\% of the total number of rings and the rest of the area is consists of pentagons (``R5''). Increase of the healing temperature to 2100 K results in significant changes: the number of ``R6'' rings grows to approximately 60\% in the cost of ``R5'' rings. Thus, that increase of the temperature results in more ``graphitic'' quality of the healed area.

Area outside of the nanopore is also affected by the healing process due to insertion healing mechanism. It results in the formation of heptagons (``R7'') in the cost of hexagons (``R6''). Unlike the structure of healed area, this process is practically unaffected by temperature, the number of R7 rings is directly proportional to the healing time.

\section{Conclusions}

The ability to control annealing and healing of graphene is essential next step in  future applications of the graphene based materials. We have found at least two healing mechanisms for graphene: one is the fully expected from previous discussions in the literature edge healing mechanism~\cite{JPC2010689, Artyukhov18092012, JPC2010689}. We also find another mechanism we call the ``atom insertion'' that  it appears, not have been discussed earlier.

We verify the natural expectation that the rate of nanopore healing is significantly affected by temperature and increasing linearly with T, at least within available MC data. Therefore, our results suggest possibility to control the final size of healed pore by terminating annealing at a prescribed time.

We also  confirm  that healing is statistical process that tends to produce high asymmetries in re-grown pores (se Fig.~\ref{fig1f}). This statistical irregularity can itself be factor that would limit the quality of healed nanopore and therefore this aspect might limit applicability of this healing method  for  potential applications that use controlled size graphene nanopore.

We have analyzed the structure of the healed area and found significant number of pentagons and heptagons being produced. Moreover, the number of pentagons and heptagons is not equal, indicating change of topology.

\section{Methods}

In our simulations we study graphene by means of molecular Monte Carlo simulations. The interatomic interactions are calculated with the LCBOPII potential~\cite{Los2005} that we have shown to describe well the elastic and thermal properties of graphene~\cite{KVZ2009, ZLKF10}. It is important to note that this potential allows bond breaking and formation with realistic energy barriers. Also it gives a significantly better description of the lattice dynamics~\cite{Karssemeijer2011} than the Tersoff potential~\cite{Tersoff1988}.

We perform MC simulations at finite temperature T with periodic boundary conditions for a sample of N = 4032 atoms (ideal sample, without nanopore) with equilibrium size at zero temperature of 103.30~\AA~in the x direction and 102.24~\AA~in the y direction. Nanopores have been created in the following way: first, atoms within the circles of pre-defined radius (from 5 to 30~\AA~with the step of 5~\AA) have been removed. Second, the sample was annealed at 4200~K to simulate the damage induced by the electronic beam at the edges of the nanopore. Finally, samples were cooled to 300~K and re-equilibrated at this temperature. Table~\ref{table:01} shows parameters of the obtained samples, with radii 5, 10, 15, 20, 25 and 30~\AA.

\begin{table}
  \caption{Parameters of the samples used in simulations}
  \label{table:01}
  \begin{tabular}{lll}
    \hline
    R,~\AA & Number of atoms removed & Atoms left \\
    \hline
    5 & 28 & 4004\\
    10 & 122 & 3910\\
    15 & 270 & 3762\\
    20 & 474 & 3558\\
    25 & 748 & 3284\\
    30 & 1080 &  2952\\
    \hline
  \end{tabular}
\end{table}

To simulate healing, carbon atoms were added randomly is small series of 9 or 16 atoms at the both sides of the sample (totally 18 or 32 atoms in each series). In both cases atoms were added at the random height from 3 to 8~\AA~below or above the sample spreaded over the area 21x21~\AA~or 40x40~\AA~for the first and second cases respectively. It was checked, that minimum distance between atoms was at least 3 \AA, thus neither two atoms can become immediately bonded between or with the sample. The atoms were added relatively  close to the graphene sample to speedup simulations. In general, atoms can be added far away from the graphene sample, but such procedure significantly increases simulation time. After atoms were added, the evolution of the system was simulated for 50000 MC steps. %This time is enough for free atom to get attached to the main sample.

 After finishing the simulation, new series of atoms were added as described above, and system were run for new 50000 steps. This cycle was repeated until the complete healing the nanopore or until pore reaches required size. After finishing the complete healing cycle, the structure of the sample was analyzed as described in {\em Structure of the healed nanopore} section of this paper.

%% classes.  Rather than use \section, an appropriate macro is
%% provided that will always work.
%%%%%%%%%%%%%%%%%%%%%%%%%%%%%%%%%%%%%%%%%%%%%%%%%%%%%%%%%%%%%%%%%%%%%
\acknowledgement
The research leading to these results has received funding from the NORDITA, DOE, VCB 621-2012-2983 and  European Research Council under the European Union's Seventh Framework Programme (FP/2007-2013) / ERC Grant Agreement n. [321031].  The computations were performed on resources provided by SNIC through Uppsala Multidisciplinary Center for Advanced Computational Science (UPPMAX) under Project p001/12-288.

%% \suppinfo macro.
%%%%%%%%%%%%%%%%%%%%%%%%%%%%%%%%%%%%%%%%%%%%%%%%%%%%%%%%%%%%%%%%%%%%%
%\suppinfo
%
%The entire \textsf{achemso} bundle is generated by running
%\texttt{achemso.dtx} through \TeX. Running \LaTeX\ on the same file
%will generate the general documentation for both the class and
%package files.

\bibliography{bib_healing}

\end{document}